\documentclass[prl,reprint, amsmath,amssymb,
 aps, superscriptaddress,floatfix
]{revtex4-2}
\usepackage{graphicx}
\usepackage{dcolumn}
\usepackage{bm}
\usepackage{wrapfig}
\usepackage{sidecap}
\usepackage{amsmath}
\usepackage{sidecap}
\usepackage{verbatim}
\usepackage{amsfonts}
\usepackage{amssymb}
\usepackage{siunitx}
\usepackage{graphicx}
\usepackage{url,plain} 
\usepackage{color}

\begin{document}

\preprint{APS/123-QED}

\title{X-ray diffraction reveals the consequences of strong deformation in thin smectic films: dilation and chevron formation\\ }
\author{Jean de Dieu Niyonzima}
 \affiliation{Sorbonne Université, CNRS, Institut des Nanosciences de Paris (INSP), F-75005 Paris, France}%
 \affiliation{Physics department, School of Science, College of Science and Technology, University of Rwanda, Po.Box: 3900 Kigali, Rwanda.}
\author{Haifa Jeridi}%
 \affiliation{ECE Research Center (LyRIDS), ECE Engineering School, 75015 Paris, France.}%
\affiliation{Sorbonne Université, CNRS, Institut des Nanosciences de Paris (INSP), F-75005 Paris, France}%
\author{Lamya  Essaoui}
 \affiliation{Sorbonne Université, CNRS, Institut des Nanosciences de Paris (INSP), F-75005 Paris, France}
\author{Caterina Tosarelli}
 \affiliation{Sorbonne Université, CNRS, Institut des Nanosciences de Paris (INSP), F-75005 Paris, France}
\author{Alina Vlad }
 \affiliation{Synchrotron SOLEIL, BP 48, L'Orme des Merisiers, 91192 Gif sur Yvette Cedex, France}
\author{Alessandro Coati}
 \affiliation{Synchrotron SOLEIL, BP 48, L'Orme des Merisiers, 91192 Gif sur Yvette Cedex, France}
   \author{Sebastien Royer}
  \affiliation{Sorbonne Université, CNRS, Institut des Nanosciences de Paris (INSP), F-75005 Paris, France}
  \author{Isabelle Trimaille}
 \affiliation{Sorbonne Université, CNRS, Institut des Nanosciences de Paris (INSP), F-75005 Paris, France} 
\author{ Michel Goldmann}
 \affiliation{Sorbonne Université, CNRS, Institut des Nanosciences de Paris (INSP), F-75005 Paris, France}
 \author{ Bruno Gallas}
 \affiliation{Sorbonne Université, CNRS, Institut des Nanosciences de Paris (INSP), F-75005 Paris, France}
 \author{ Doru Constantin}
 \affiliation{%
Université de Strasbourg, CNRS, Institut Charles, 67034 Strasbourg Cedex, France.}%
 \author{ David Babonneau}
\affiliation{%
 Université de Poitiers, ISAE-ENSMA, CNRS, PPRIME, Poitiers, France}%
  \author{Yves Garreau}
  \affiliation{Synchrotron SOLEIL, BP 48, L'Orme des Merisiers, 91192 Gif sur Yvette Cedex, France}
  \affiliation{Université Paris Cité, CNRS, Laboratoire Matériaux et Phénomènes Quantiques, F-75013, Paris, France.}
\author{Bernard Croset}
\affiliation{Sorbonne Université, CNRS, Institut des Nanosciences de Paris (INSP), F-75005 Paris, France}%
\author{Samo Kralj}
\affiliation{Department of Physics, Faculty of Natural Sciences and Mathematics, University of Maribor, 2000 Maribor, Slovenia}
\affiliation{Condensed Matter Physics Department, Jožef Stefan Institute, 1000 Ljubljana, Slovenia}
\author{Randall D. Kamien}
\affiliation{Department of Physics and Astronomy, University of Pennsylvania, Philadelphia, PA 19104, USA}
 \author{Emmanuelle Lacaze}
 \email{ emmanuelle.lacaze@insp.jussieu.fr}
 \affiliation{Sorbonne Université, CNRS, Institut des Nanosciences de Paris (INSP), F-75005 Paris, France}%

\date{\today}

\begin{abstract}
Smectic liquid crystals can be viewed as model systems for lamellar structures for which there has been extensive theoretical development. We demonstrate that a nonlinear energy description is required with respect to the usual Landau-de Gennes elasticity in order to explain the observed layer spacing of highly curved smectic layers. Using X-ray diffraction we have quantitatively determined the dilation of bent layers distorted by antagonistic anchoring (as high as 1.8\% of dilation for the most bent smectic layers) and accurately described it by the minimal nonlinear expression for energy. We observe a 1$^{\circ}$ tilt of planar layers that are connected to the curved layers. This value is consistent with simple energetic calculations, demonstrating how the bending energy impacts the overall structure of a thin distorted smectic film. Finally, we show that combined X-ray measurements and theoretical modeling allow for the quantitative determination of the number of curved smectic layers and of the resulting thickness of the dilated region with unprecedented precision.

 \end{abstract}

\maketitle


The smectic state of liquid crystals bridges the study of broken orientational symmetry {\sl \`a la} nematic order, the statistical mechanics of membranes, and the long range periodic order in crystals.  As such, it can be probed optically, mechanically, and through X-ray diffraction.  From a theoretical perspective, it provides an arena to systematically study the effects of nonlinear elasticity via fluctuations \cite{grinpel}, topological defects (where the distortions can be large) \cite{LavIsh,brener,santangelo}, and, as we will demonstrate here, even in complexions arising from antagonistic boundary conditions.  By focusing on regions of high layer curvature using X-ray diffraction we have measured the layer spacing with high resolution in the wave-vector transfer and find that it is accurately described by the nonlinear response of bent layers.  As we will discuss, the nonlinearities we include are required by symmetry and do not require any additional elastic constants providing a parsimonious explanation of the data. Thus we demonstrate, even in a static configuration, the essential role of rotational invariance in systems with spontaneously broken symmetries.

\begin{figure} \includegraphics[width=0.5\textwidth]{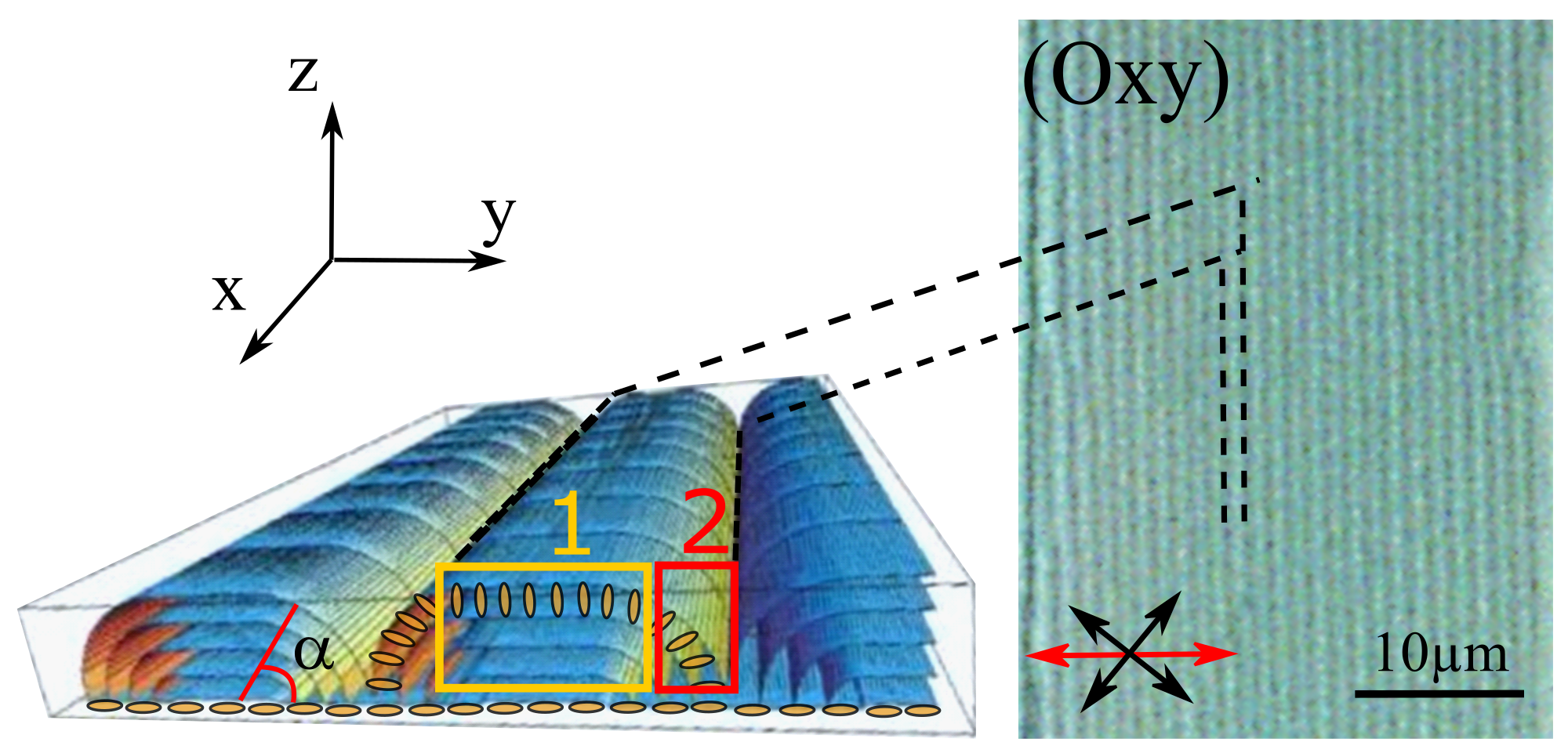}
\caption{Optical Microscopy image in the reflection mode between crossed polarizers providing a top view of the 8CB film. The red arrow shows the direction of PVA rubbing, and the black crosses show the polarizer directions. Flattened hemicylinders are schematized on the left. Zone 1 (yellow) corresponds to the central part of the flattened hemicylinders. Zone 2 (red) corresponds to the curved layers at the edges of the flattened hemicylinders. The 8CB molecules are schematized at the interfaces in yellow.}
\label{fig1}
\end{figure}

We deposited a smectic film of 4-\textit {n}-octyl-4'-cyanobiphenyl (8CB) of thickness 180 nm on a rubbed polyvinyl alcohol (PVA) coated glass surface in contact with air, the former providing planar anchoring and the latter homeotropic anchoring of the nematic director. The resulting structure is made up of unidirectional stripes of a period around 600 nm perpendicular to the rubbing direction, as observed by optical microscopy (Fig. \ref{fig1}).  By combining grazing incidence and transmission X-ray diffraction with ellipsometry we have previously determined that the resulting configuration is similar, in spirit, to an oily streak texture \cite{michel2004, coursault2016self}.  Due to the strong planar unidirectional anchoring induced by the rubbed PVA substrate, perpendicular smectic layers are present throughout, to a thickness of around 20-30 nm. Flattened hemicylinders are formed on top of these perpendicular layers to oblige the strong homeotropic anchoring at the air interface. A sharp, topological grain boundary is thus induced in the flattened hemicylinder between the perpendicular smectic layers below and the planar smectic layers above (below zone 1 on Fig. \ref{fig1}) \cite{coursault2016self}. Curved smectic layers form at the two edges of the flattened hemicylinders (zone 2 on Fig. \ref{fig1}) \cite{michel2004}.

\begin{figure*}
    \centering
\includegraphics[width=\linewidth]{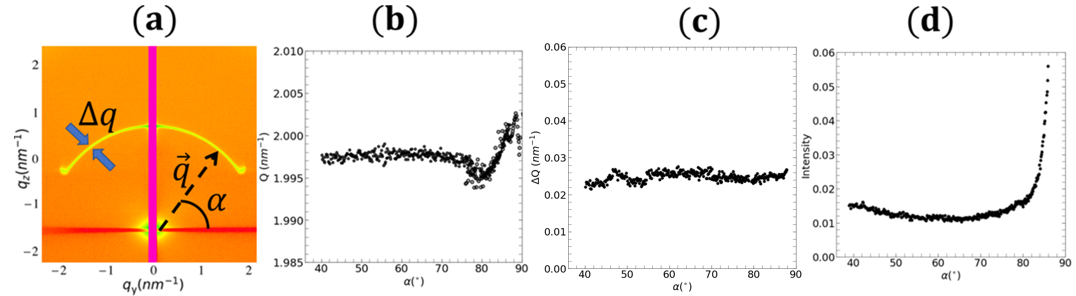}
    \caption{(a) The intensity scattered by smectic layers in the flattened hemicylinders for an incident angle 0.6$^{\circ}$.  (b) The position of intensity maximum,  $Q$ as a function of $\alpha$. (c) The ring width $\Delta Q$. (d) The integrated intensity $I$.}  
    \label{fig2}
\end{figure*}

By X-ray diffraction under grazing incidence conditions (see section {\sl End Matter}) we reveal a ring-like scattering signal associated with these continuously rotating smectic layers at the edges of the flattened hemicylinders (Fig. \ref{fig2}(a)) \cite{coursault2016self}. For a given position in the scattering ring at an angle $\alpha$ corresponding to the direction of the wave-vector transfer $\textbf{q}$ (Fig. \ref{fig2}(a)), the signal corresponds to those smectic layers with their normal parallel to $\textbf{q}$ (Fig. \ref{fig1}). To ensure that the Bragg condition is satisfied, the incident angle is chosen to be the Bragg angle $\arcsin(\lambda_o/2d) \sim 0.6^{\circ}$, where $d=3.14$ nm is the 8CB smectic spacing and $\lambda_o = 0.067$ nm is the X-ray wavelength \cite{leadbetter1979, clark_x-ray_1993}. However, this incident angle value obscures the signal from the curved smectic layers below $\alpha = 40^\circ$ associated with the structure at the base of the flattened hemicylinders, closest to the underlying perpendicular layers.
We will thus focus on the structure of the curved smectic layers rotating from $\alpha = 40^\circ$ (zone 2 in Fig. \ref{fig1}) and on their connection with the central planar layers (zone 1 in Fig. \ref{fig1}) that occurs close to $\alpha = 90^\circ$.

\begin{figure}
    \centering
\includegraphics[width=1\linewidth]{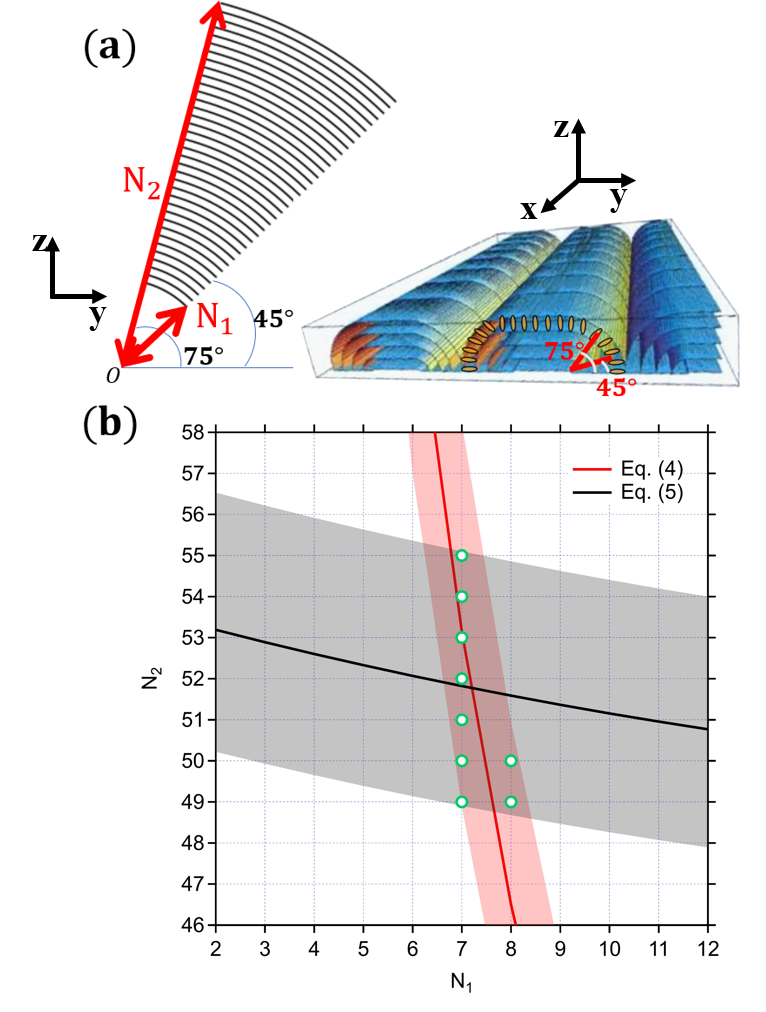}
   \caption{ (a) Slice of quarter cylinder between $\alpha = 45^\circ$ and $\alpha = 75^\circ$ with $N_{1}$ and $N_{2}$ shown. For the sake of clarity, only 29 layers are shown (b) Curves of $N_{2}$ as a function of $N_{1}$. In red it corresponds to  (\ref{eq:integ}) with the influence of the uncertainty in $(Q - Q_0)$ also shown by the red shaded zone. In black it corresponds to (\ref{eq:I_Q}) with the influence of the uncertainty in $\Delta Q $ also shown by the black shaded zone. The green dots correspond to the corresponding possible ($N_1$, $N_2$) values.}
    \label{fig3}
\end{figure}

In Fig. \ref{fig2}(b) we show the absolute value of $\textbf{q}$, $Q$, measured as the position of maximum intensity in the scattering ring as a function of $\alpha$. Fig. \ref{fig2}(c)
shows the ring width, $\Delta Q$, and Fig. \ref{fig2}(d) shows the corresponding integrated intensity, $I$. We focus on only one side of the flattened hemicylinders ($40^\circ \leq \alpha \leq 90^\circ$). This is due to the symmetry of the data observed on the scattering ring shown in Fig. \ref{fig2}(a) which corresponds to symmetrical flattened hemicylinders. For a given $\alpha$, $Q$ is inversely proportional to the average spacing between the layers ($Q = 2\pi /d$) when the Bragg condition is fulfilled. Both $I$ and $\Delta Q$ depend on the number of smectic layers associated with $\alpha$, as we will see hereafter for $\Delta Q$. The nearly constant values of $I$, $Q$, and $\Delta Q$ between $\alpha = 45^\circ$ and $\alpha = 75^\circ$ (Fig. \ref{fig2}) indicate in this $\alpha$ range an almost constant number of smectic layers rotating around their curvature center with constant average spacing $d$ (Fig. \ref{fig3}(a)).
At $\alpha\approx90^{\circ}$, associated with the central planar layers (zone 1 in Fig. \ref{fig1}), $Q = 2.003$ nm\textsuperscript{-1} (Fig. \ref{fig2}(b)). Being confined between air and the topological grain boundary the central layers don't experience any modification, neither from elasticity nor from confining surfaces. We expect the layer spacing of the central layers to be that of a bulk 8CB smectic, $d_0$. The value of $Q_{0}$ of 2.003 nm\textsuperscript{-1} is associated with $d_0=3.14$ nm and accordingly close to the already published values \cite{leadbetter1979, clark_x-ray_1993} if we take into account the experimental precision of maximum value 0.015 nm$^{-1}$ that may only affect the absolute value of $Q$, not its relative variation. Between $\alpha=45^{\circ}$ and $\alpha=75^{\circ}$ (zone 2 in Fig. \ref{fig1} - Fig. \ref{fig3}) $Q$ is constant, but smaller, with $Q_0-Q=0.00537\pm0.00047$ nm$^{-1}$ (or $(Q_0-Q)/Q=0.00269\pm0.00023$). The small but clear decrease of $Q$ implies that the curved layers in most of the quarter cylinder are dilated compared to the central planar layers (the layer spacing $d = 2\pi /Q$ is larger between $\alpha = 45^\circ$ and $\alpha = 75^\circ$ than around $\alpha \approx 90^\circ$). 

The origin of this dilation is due to the small radius of curvature imposed on the curved layers of these thin smectic films.  To reduce the high bending energy near the center of the quarter cylinders, the layers will dilate to decrease the curvature.  In order to properly model the elastic response it is essential that any elastic free energy we use must remain valid even for large deformations of layer spacing and orientation.  This necessarily requires that the expressions for strain and curvature be extended to account for rotational invariance.  In particular, this leads to a nonlinear relation between strain and the phonon mode representing the displacement of the layers \cite{grinpel}.   Subsequently, we can develop the free energy as a series expansion in the strain and the curvature, beginning at harmonic ({\sl i.e.}, quadratic) order.  We could include {\sl anharmonic} terms with unknown moduli.  We will see, however, that we can interpret our results quantitatively just by using a harmonic theory with known elastic moduli but with nonlinear strain.  Thus we need not invoke any new elastic coupling constants and rely entirely on symmetry to set the relative coefficients of the higher order terms.
To this end, we have quantitatively calculated the dilation for each smectic layer through a minimization of the non-linear elastic energy \cite{AG}
\begin{equation}
F = {\frac{B}{2}}\int d^2 x \left\{{\frac{1}{4}}[(\nabla\phi)^2-1]^2 + {\lambda^2}(\nabla^2\phi)^2\right\}
\label{eq:ag}
\end{equation}
where $B$ is the bulk modulus, $\lambda=\sqrt{K/B}$ is the penetration length ($K$ the layer bend elastic constant) and $\phi$ is the phase variable of the smectic density wave, {\sl i.e.,} $\rho({\bf x}) = \rho_0 + \rho_1\cos[2\pi\phi({\bf x})/d_0]$.  The first term in (\ref{eq:ag}) appears explicitly to measure compression/dilation energy while the second term measures curvature: both are harmonic in strain and curvature.  We can see explicitly how the nonlinear realization of the symmetry appears in (\ref{eq:ag}): there are both quadratic and quartic terms in $\nabla\phi$ but their relative co\"efficient {\sl must} be chosen to set the layer spacing to $d_0$.  It is precisely this requirement that leads to the celebrated anomalous fluctuations of the smectic layers \cite{grinpel}.   Energy extremals satisfy the (fourth order) Euler-Lagrange equation
\begin{equation}\label{eq:el}
\nabla\cdot\left(\nabla\phi\left[(\nabla\phi)^2-1\right]\right) = {2\lambda^2}\nabla^2\nabla^2\phi
\end{equation}
Note that when layers are equally spaced, the left-hand side vanishes since $(\nabla\phi)^2=1$.  However, in the quarter cylinders, the right-hand side does not vanish.  To see this, we use the radial polar co\"ordinate $r$ in the $xz$ cross section: $\phi=r$ and so $\lambda^2\nabla^2\nabla^2\phi = \lambda^2/r^3$.  Because $\lambda$ is comparable to the molecular length (i.e., $\lambda \sim d_0$), we see that the curvature term is suppressed by a factor of $(\lambda/r)^2$ when we are multiple layers away from center of curvature.  In this regime we can solve (\ref{eq:el}) perturbatively.   Letting $\phi=r+ \Delta(r)$ (\ref{eq:el}) becomes, to linear order in $\Delta$:
\begin{equation}
(\Delta'' + {\frac{\Delta'}{r}}) = {\frac{\lambda^2}{r^3}}
\end{equation}
from which it follows that $\Delta(r)=r_0+\lambda^2/r$.  The radius of the $N^{\hbox{\sl th}}$ layer found via $\phi=Nd_0$, is for large $N$, $r_N \approx r_0 + d_0(N-1/N)$ where $r_0$ is an unknown parameter that sets the location of the initial layer.  In future work, we will explore the physics that leads to this small but measurable offset. Not only is this solution consistent with the perturbative solution, {\sl i.e.,} $\Delta'/\phi' \sim \lambda^2/r^2$, it also shows that the layer spacing for the $N^{\hbox{\sl th}}$ layer is $d_N=r_{N+1} - r_{N}= d_0(1 +  \frac{1}{N(N+1)})$.  Because $(d_0 - d_N)/d_0 < 0$ the layers expand in response to being bent. 

We parameterize the layers of the quarter cylinder by $N_1$ and $N_2$:  $N_{2}$ is the number of the last layer of largest curvature radius and $N_{1}$ the one of the first layer, of the smallest curvature radius as shown in Fig. \ref{fig3}(a). We expect that layers with radius smaller than $N_1 d_0$ will become too costly near the center of curvature due to the diverging elastic energy. These missing curved layers may be replaced either by disordered liquid crystal (nematic or isotropic area), by lengthened central smectic layers close to the center of curvature, or by both \cite{michel2006, coursault2016self}.
Using our perturbative solution, we  calculate the average dilation per layer $d = {\frac{r_{N_2} - r_{N_1}}{N_2-N_1}}$ as
\begin{eqnarray}
\label{eq:integ}
{\frac{d - d_0}{d_0}} = {\frac{Q_0 - Q}{Q}} = {\frac{1}{(N_1 N_2)}}
\end{eqnarray}
$(N_2-N_1)d_0$ is the thickness of the quarter cylinder we would find were there no curvature energy ({\sl i.e.}, when $\lambda=0$).

To ascertain $N_1$ and $N_2$, we turn to the data. As stated earlier, the measured dilation of the curved layers $(Q_0-Q)/Q=0.00269\pm0.00023$ (Fig. \ref{fig2}(b)) allows us to draw a first curve that relates $N_{1}$ to $N_{2}$ (in red in Fig. \ref{fig3}(b)).

Alternatively, $N_1$ and $N_2$ can be calculated from the evolution of $\Delta Q$ for a quarter cylinder as a function of $N_{2}$ and $N_{1}$. As expected $1/\Delta Q$ linearly depend on the number of layers $N_{2} - N_{1}$, (see End Matter - Fig. \ref{figureSM2}):

\begin{equation}
\frac{1}{\Delta Q} = c(N_{1}) \times (N_{2} - N_{1})  + d(N_{1})
\label{eq:I_Q}
\end{equation}

where $c(N_{1})$ and $d(N_{1})$  have been calculated (see End Matter - Fig. \ref{figureSM2}).

We use the experimental measurement of $\Delta Q = 0.0249 \pm 0.0014$ nm$^{-1}$ (Fig. \ref{fig2}(c)) to obtain a new curve relating $N_{1}$ and $N_{2}$ (black curve in Fig. \ref{fig3}(b)). The intersection of the red and black curves in Fig. \ref{fig3}(b) allows us to estimate the appropriate range of ($N_{1}$, $N_{2}$) values. They are shown as green dots on Fig. \ref{fig3}(b). We obtain $N_{1} = 7$ with $N_{2} = 52 \pm 3$ or $N_{1} = 8$ with $N_{2} = 49$ or $N_{2} = 50$.
$N_{2} = 52$ corresponds to a quarter cylinder radius of 163nm. The perpendicular layers that form all along the substrate have an estimated thickness of around 20 nm \cite{coursault2016self}. Adding this 20 nm base layer to the quarter cylinder yields an overall thickness very close to the expected 180 nm.
These findings confirm the overall picture of the above model: the curved layers display dilation to allow for a slight increase of their radius of curvature. The dilation $(d_N - d_0)/d_0$ evolves as $1/N(N+1)$ with $N$ the number of the considered layer counted from the curvature center. When the first complete layer is the 7$^{\hbox{\sl th}}$ (or the 8$^{\hbox{\sl th}}$), dilations as high as 1.8\%  can be reached. Dilation in response to large curvature of the smectic layers is a general feature:
it certainly arises in focal conics for those smectic layers close to the elliptic and hyperbolic focal lines. Here, it is experimentally revealed for two reasons: we focus on thin films that allow us to enhance the signal of the most distorted smectic layers and, on the other hand, the high $Q$ resolution allowing for measurements of average interlayer spacings with relative variations of only 0.25\%. This experimental demonstration that non-negligible dilation occurs for the most distorted smectic layers is finally one of the rare experimental demonstrations of the non-linear elasticity of smectic materials at length scales as small as ten times the molecular size.
 \begin{figure}
    \centering
\includegraphics[width=1\linewidth]{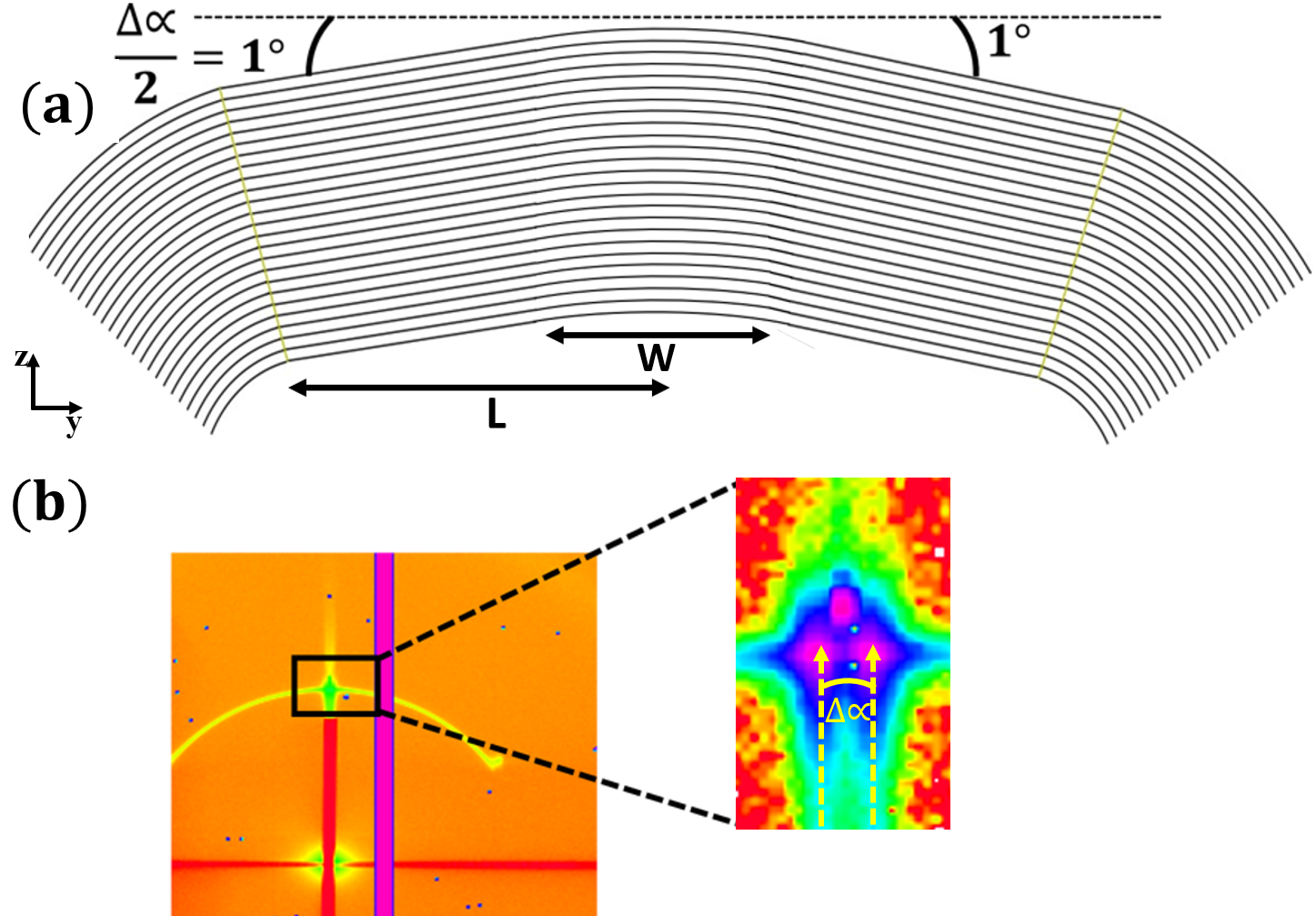}
   \caption{(a) Internal structure of the hemicylinder with the curved smectic layers at the edges of the hemicylinder, the central inclined and straight smectic layers, the central chevron of width $W$. For the sake of clarity, only 23 layers are shown. The junction between the curved layers at the edges and the central ones is shown by the yellow inclined line. (b) zoom on the top of the scattering ring for $\alpha$ values close to $90^\circ$. An incident angle of 0.61$^{\circ}$ has been used to allow the substrate reflectivity at exactly $\alpha=90^\circ$ to be outside the scattering ring. Along the scattering ring only two peaks appear at 89$^{\circ}$ and 91$^{\circ}$ thus separated by $\Delta \alpha = 2^\circ$.}
    \label{fig4}
\end{figure}
The large bending energy of the smectic curved layers not only induces dilation of these curved layers but also impacts the whole structure of these thin distorted smectic films. We expect the central layers of the hemicylinder to be flat in order to induce the smallest possible surface energy. Accordingly the intensity is strongly increased around $\alpha = 90^\circ$ (Fig. \ref{fig2}(d) - \ref{fig4}(b)). However, if we focus precisely on the upper part of the scattering ring, we observe {\sl two} symmetric maxima at $\alpha = 89^\circ$ and $\alpha = 91^\circ$ (Fig. \ref{fig4}(b)). They demonstrate that the central smectic layers are {\sl not} strictly flat: two symmetrical zones made of straight but tilted smectic layers are formed that lead to a central chevron of width $W$ that joins the straight and tilted smectic layers with a dihedral angle of $\Delta \alpha = 2^\circ$ (Fig. \ref{fig4}(a)).
The mismatch in the spacing between the central planar layers and the dilated curved layers in the quarter cylinder forces an angle between their two normals.  However, a tilted slightly curved boundary between the two regions can accommodate strictly flat planar layers in the central region.  So why do the central layers tilt when this increases the surface energy between the mesogens and the air?  It can be understood through the observation that a $\Delta \alpha /2$ tilt of the central layers induces a reduction in the arclength of the curved layers in the quarter cylinder and a concomitant reduction in the bend and compression energy.  This is, of course, balanced against the extra area of the tilted 
central region.  If we excise a $\Delta \alpha /2$ wedge the saved bending energy is $E_{b} = - {\frac{1}{2}}K\ln(N_2/ N_1) \Delta \alpha/2$ while the corresponding dilation energy is two orders of magnitude smaller and neglected.  
The bent region in the center has a low energy, $E_{\rm central} = 2(K/\lambda) [\tan(\Delta \alpha /2) - (\Delta \alpha /2))]\cos(\Delta \alpha /2)$ \cite{Blanc1999}. It is also two orders of magnitude smaller than the bending energy of the quarter cylinder region. Taking into account the known width, $W = \lambda /(\Delta \alpha /2)$, of a central chevron of small angle $\Delta \alpha$ (Fig. \ref{fig4}(a)) \cite{DeGennes} the increased surface energy due to the tilted central layers is $E_{s} \approx \gamma ({\frac{1}{2}}L(\Delta \alpha /2)^{2} - {\frac{1}{6}}\lambda (\Delta \alpha /2))$, where $\gamma$ is the 8CB surface tension and $L$ is the central lateral size of the hemicylinders (Fig. \ref{fig4}(a)). The minimization of the remaining two energy terms leads to $\Delta\alpha/2 \sim {\frac{K\ln(N_2/ N_1)}{(2L \gamma)}} + (\lambda/6L)$. With the experimental value $L=185$ nm \cite{coursault2016self}, the surface tension of 8CB  $\gamma = 30$ mJ/m$^{2}$ \cite{tintaru2001, schuring2001}, $N_1=7$, $N_2=52$ we find that for $\Delta \alpha /2 = 1^\circ$ $K = 8.09 \times 10^{-11}$ N: the smectic 8CB elastic constants have been measured  with values varying between $K = 5.2 \times 10^{-12}$ N and $K = 1.8 \times 10^{-10}$ N \cite{bradshaw1985, Benzekri1992, Zywocinski2000} and thus our estimate is compatible with prior art. We thus conclude that the observed tilt is compatible with the known properties of 8CB and that it is driven by the necessity of decreasing the bending energy of the curved smectic layers.

In conclusion, by combining synchrotron-based X-ray diffraction with a theoretical energetic model we demonstrate rare consequences of the non-linear character of the smectic elastic energy. In order to reduce the bending energy dilation is induced in smectic layers: $(d_N - d_0)/d_0 \sim 1/N(N+1)$, where $N$ labels the smectic layer counted from the curvature center. 
It is shown by studying thin smectic 8CB films in which the distortion is managed by antagonistic anchoring at the two interfaces (planar unidirectional and homeotropic). However, this is a general behavior that does not depend on the specific elastic parameters of the material and cannot be described by classical Landau-de Gennes elasticity. It should exist for all curved smectic layers close to their curvature center. For example, this will occur near the focal lines in a focal conic and will occur in any material with a lamellar ground state. In the arrays of flattened hemicylinders that appear in these distorted 8CB films we have shown quantitative dilations as large as 1.8\%.  In fact, the bending energy associated with curvature radii 7-50 times the layer spacing can successfully compete with the surface energy of a large central region, leading to a non-negligible tilt of the straight smectic layers attached to the curved layers in the flattened hemicylinders.  We also show that combined synchrotron-based X-ray diffraction and energetic consideration allow to precisely determine (within 10\%) the overall structure of these thin distorted smectic films despite the corresponding small thickness of 180nm.
\vskip.3truein
\centerline{\bf End Matter}

\begin{figure}
\centering
\includegraphics[width=0.8\linewidth]{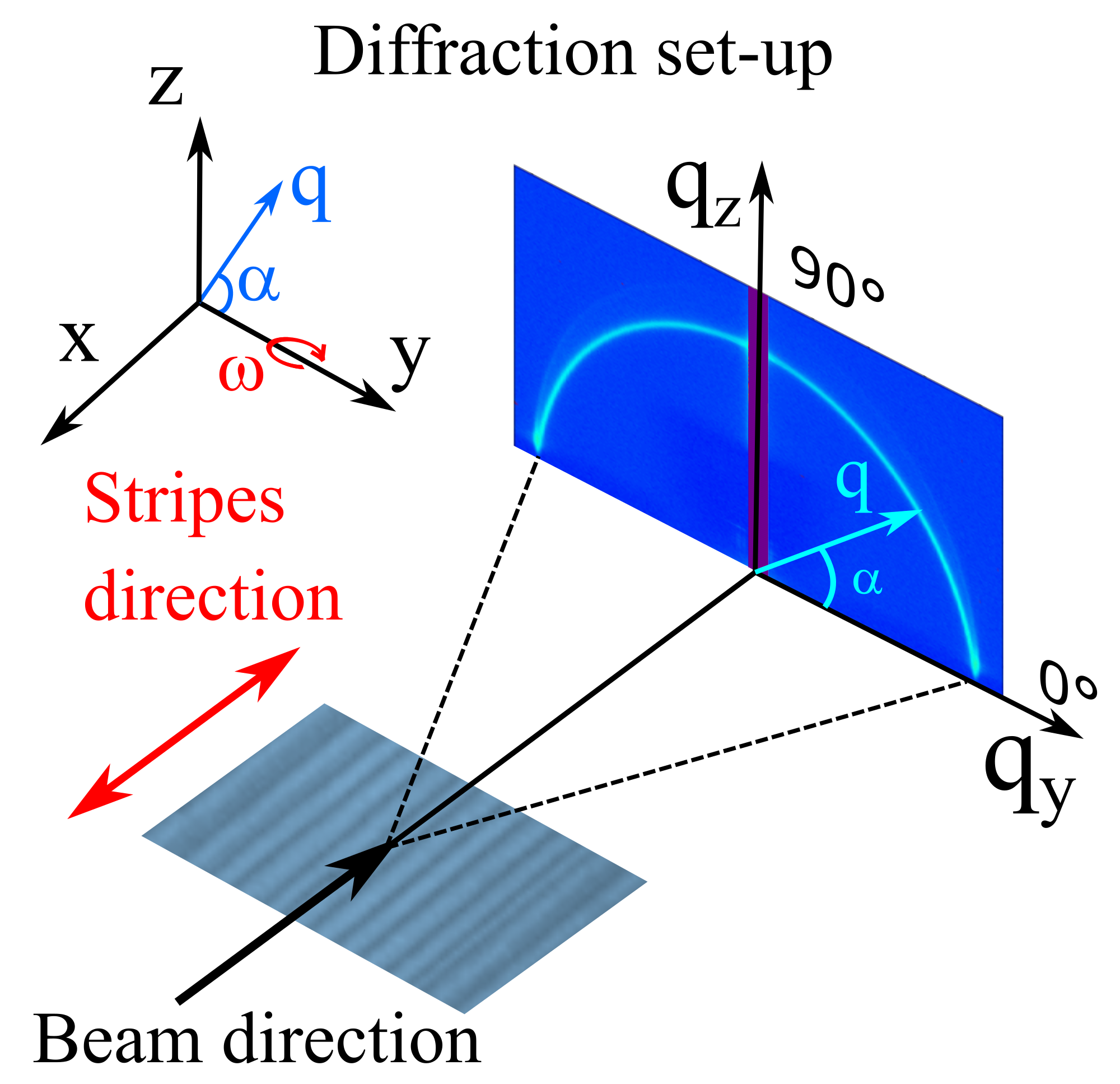}
\caption{X-ray diffraction setup, where the smectic stripes are almost parallel to the beam showing the ring-like scattered signal on the Eiger detector.}
\label{figureSM1}
\end{figure}
\noindent\textit{Sample preparation}\newline
Smectic films with an average thickness of 180nm were prepared by depositing smectic 4-n-octyl-4’-cyanobiphenyl (8CB, from Sigma Aldrich) onto rubbed polyvinyl alcohol (PVA, from Sigma Aldrich). The PVA layer, approximately 10nm thick, was prepared by spin-coating a 100 $\mu$L droplet of a 0.5 wt\% aqueous solution of PVA, with an acceleration of 400 rpm/s and a speed of 3000 rpm for 30 seconds on glass slides previously cleaned with precision (thickness 130 $\mu$m).
Spin-coating was also employed for the deposition of 8CB once the PVA layer had been rubbed with a rubbing machine using a roller covered by velvet tissue. The 8CB was dissolved in toluene (0.2 M). Subsequently, a 50$\mu$L droplet of this solution was deposited onto the rubbed PVA layer and spin-coated at a speed of 3000 rpm with an acceleration ranging between 500 rpm/s  and 1000 rpm/s for 30 seconds. The average thickness has been determined through a careful mapping of the overall sample by optical microscopy. Between parallel polarizers, the local colors are associated with optical phenomena induced by the interfaces of the LC film. The film thickness is thus extracted through the analysis of the color bands as seen through polarized microscopy. We have established a corresponding color chart by using an average optical index $n=(n_e+2n_o)/3$, with 8CB extraordinary index $n_e=1.67$ and ordinary index $n_o=1.52$. 
\newline
\noindent\textit{X-ray scattering measurements}\newline
The X-ray diffraction measurements were performed at the SIXS beamline of the SOLEIL synchrotron facility. The photon energy was fixed at 18.44 keV, and the X-ray beam size was set at 300 $\mu$m $\times$ 100 $\mu$m. The scattering signal was collected on a 2D Eiger 1M hybrid pixel detector (DECTRIS), which is located 1700 mm from the sample. The experimental setup is described in Fig. \ref{figureSM1}, where the observed stripes from optical microscopy (Fig. \ref{fig1}) are almost parallel to the incident beam. The X-ray beam probes an area with dimensions of 300 $\mu$m wide and 18 mm long, matching the entire sample length. For a given $\alpha$ the scattering ring was fit to a Gaussian curve allowing us to extract $Q$, the position of maximum intensity in the scattering ring, $\Delta Q$, the full width at half maximum (FWHM) as well as  $I$ the integrated intensity (Fig. \ref{fig2}). Between $\alpha = 45^\circ$ and $\alpha = 75^\circ$, we have measured the average $Q$ with its standard deviation and the average $\Delta Q$ with its standard deviation leading to $Q=1.99763\pm0.00047$ nm\textsuperscript{-1} and $\Delta Q = 0.0249 \pm 0.0014$ nm\textsuperscript{-1}
\newline
\noindent\textit{Cylindrical calculation of a quarter cylinder}\newline
We have calculated the scattering from a complete quarter cylinder for a given wave-vector transfer $\textbf{q}$ of orientation $\alpha$ and of absolute value $Q$ :
\begin{equation}\label{Amplitude1a}
     A (Q,r,\alpha) =  L \int^{\pi/2}_{ 0} d\beta \int^{\infty}_{0} rdr\;\sum^{N_{2}}_{N_{1}} e^{iQr\cos{(\alpha - \beta)}} \delta(r-nd) 
\end{equation}
We have shown that $A(Q,r,\alpha)$ does not depend on $\alpha$, as expected, and only depends on the value of $N_{1}$ and $N_{2}$. To calculate the dependency of $\Delta Q$ as a function of $N_{1}$ and $N_{2}$, we first calculated the intensity $I = \mid A \mid^{2}$. For given values of $N_{1}$ and $N_{2}$ we have drawn the variation of $I$ as a function of $Q$. After fitting the calculated curve with a Gaussian curve, we extracted $\Delta Q$. This calculation was made for arbitrary $N_{1}$ and $N_{2}$ (Fig. \ref{figureSM2}), and we find that $\Delta Q$ can be written as in (\ref{eq:I_Q}), where 
\begin{equation}
\left\{ \begin{aligned}  
c(N_{1})  &= -3.7\times10^{-5}N^{2}_{1} +  3\times10^{-3}N_{1} + 0.738 \\
d(N_{1}) &= 0.828N_{1} + 0.396
\end{aligned} \right.
\end{equation}
\begin{figure}
\centering
\includegraphics[width=1\linewidth]{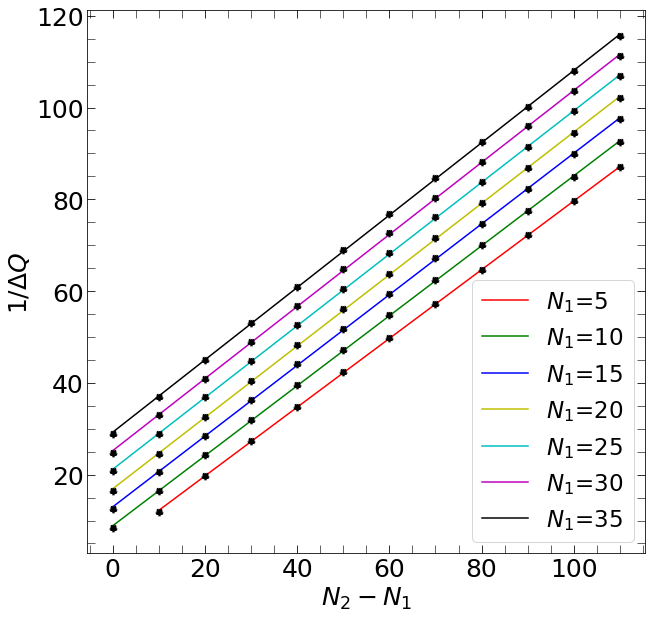}
\caption{Evolution of the inverse of the $\Delta Q$ as function of $N_{2}-N_{1}$ for different values of $N_{1}$ for a perfect quarter cylinder.}
\label{figureSM2}
\end{figure}

\begin{acknowledgments}
JdD. N. thanks the French embassy in Rwanda for having provided the PhD grant. R.D.K. thanks the Institute for Theoretical Physics at Utrecht University for their hospitality when this work was initiated. R.D.K. was supported by a Simons Investigator Grant from the Simons Foundation.
\end{acknowledgments}

\bibliography{aapmsampb}

\providecommand{\noopsort}[1]{}\providecommand{\singleletter}[1]{#1}%
\begin{thebibliography}{17}%
\makeatletter
\providecommand \@ifxundefined [1]{%
 \@ifx{#1\undefined}
}%
\providecommand \@ifnum [1]{%
 \ifnum #1\expandafter \@firstoftwo
 \else \expandafter \@secondoftwo
 \fi
}%
\providecommand \@ifx [1]{%
 \ifx #1\expandafter \@firstoftwo
 \else \expandafter \@secondoftwo
 \fi
}%
\providecommand \natexlab [1]{#1}%
\providecommand \enquote  [1]{``#1''}%
\providecommand \bibnamefont  [1]{#1}%
\providecommand \bibfnamefont [1]{#1}%
\providecommand \citenamefont [1]{#1}%
\providecommand \href@noop [0]{\@secondoftwo}%
\providecommand \href [0]{\begingroup \@sanitize@url \@href}%
\providecommand \@href[1]{\@@startlink{#1}\@@href}%
\providecommand \@@href[1]{\endgroup#1\@@endlink}%
\providecommand \@sanitize@url [0]{\catcode `\\12\catcode `\$12\catcode `\&12\catcode `\#12\catcode `\^12\catcode `\_12\catcode `\%12\relax}%
\providecommand \@@startlink[1]{}%
\providecommand \@@endlink[0]{}%
\providecommand \url  [0]{\begingroup\@sanitize@url \@url }%
\providecommand \@url [1]{\endgroup\@href {#1}{\urlprefix }}%
\providecommand \urlprefix  [0]{URL }%
\providecommand \Eprint [0]{\href }%
\providecommand \doibase [0]{https://doi.org/}%
\providecommand \selectlanguage [0]{\@gobble}%
\providecommand \bibinfo  [0]{\@secondoftwo}%
\providecommand \bibfield  [0]{\@secondoftwo}%
\providecommand \translation [1]{[#1]}%
\providecommand \BibitemOpen [0]{}%
\providecommand \bibitemStop [0]{}%
\providecommand \bibitemNoStop [0]{.\EOS\space}%
\providecommand \EOS [0]{\spacefactor3000\relax}%
\providecommand \BibitemShut  [1]{\csname bibitem#1\endcsname}%
\let\auto@bib@innerbib\@empty
\bibitem [{\citenamefont {Grinstein}\ and\ \citenamefont {Pelcovits}(1982)}]{grinpel}%
  \BibitemOpen
  \bibfield  {author} {\bibinfo {author} {\bibfnamefont {G.}~\bibnamefont {Grinstein}}\ and\ \bibinfo {author} {\bibfnamefont {R.~A.}\ \bibnamefont {Pelcovits}},\ }\bibfield  {title} {\bibinfo {title} {Nonlinear elastic theory of smectic liquid crystals},\ }\href {https://doi.org/10.1103/PhysRevA.26.915} {\bibfield  {journal} {\bibinfo  {journal} {Phys. Rev. A}\ }\textbf {\bibinfo {volume} {26}},\ \bibinfo {pages} {915} (\bibinfo {year} {1982})}\BibitemShut {NoStop}%
\bibitem [{\citenamefont {Ishikawa}\ and\ \citenamefont {Lavrentovich}(1999)}]{LavIsh}%
  \BibitemOpen
  \bibfield  {author} {\bibinfo {author} {\bibfnamefont {T.}~\bibnamefont {Ishikawa}}\ and\ \bibinfo {author} {\bibfnamefont {O.~D.}\ \bibnamefont {Lavrentovich}},\ }\bibfield  {title} {\bibinfo {title} {Dislocation profile in cholesteric finger texture},\ }\href {https://doi.org/10.1103/PhysRevE.60.R5037} {\bibfield  {journal} {\bibinfo  {journal} {Phys. Rev. E}\ }\textbf {\bibinfo {volume} {60}},\ \bibinfo {pages} {R5037} (\bibinfo {year} {1999})}\BibitemShut {NoStop}%
\bibitem [{\citenamefont {Brener}\ and\ \citenamefont {Marchenko}(1999)}]{brener}%
  \BibitemOpen
  \bibfield  {author} {\bibinfo {author} {\bibfnamefont {E.~A.}\ \bibnamefont {Brener}}\ and\ \bibinfo {author} {\bibfnamefont {V.~I.}\ \bibnamefont {Marchenko}},\ }\bibfield  {title} {\bibinfo {title} {Nonlinear theory of dislocations in smectic crystals: An exact solution},\ }\href {https://doi.org/10.1103/PhysRevE.59.R4752} {\bibfield  {journal} {\bibinfo  {journal} {Phys. Rev. E}\ }\textbf {\bibinfo {volume} {59}},\ \bibinfo {pages} {R4752} (\bibinfo {year} {1999})}\BibitemShut {NoStop}%
\bibitem [{\citenamefont {Santangelo}\ and\ \citenamefont {Kamien}(2003)}]{santangelo}%
  \BibitemOpen
  \bibfield  {author} {\bibinfo {author} {\bibfnamefont {C.~D.}\ \bibnamefont {Santangelo}}\ and\ \bibinfo {author} {\bibfnamefont {R.~D.}\ \bibnamefont {Kamien}},\ }\bibfield  {title} {\bibinfo {title} {Bogomol'nyi, prasad, and sommerfield configurations in smectics},\ }\href {https://doi.org/10.1103/PhysRevLett.91.045506} {\bibfield  {journal} {\bibinfo  {journal} {Phys. Rev. Lett.}\ }\textbf {\bibinfo {volume} {91}},\ \bibinfo {pages} {045506} (\bibinfo {year} {2003})}\BibitemShut {NoStop}%
\bibitem [{\citenamefont {Michel}\ \emph {et~al.}(2004)\citenamefont {Michel}, \citenamefont {Lacaze}, \citenamefont {Alba}, \citenamefont {De~Boissieu}, \citenamefont {Gailhanou},\ and\ \citenamefont {Goldmann}}]{michel2004}%
  \BibitemOpen
  \bibfield  {author} {\bibinfo {author} {\bibfnamefont {J.-P.}\ \bibnamefont {Michel}}, \bibinfo {author} {\bibfnamefont {E.}~\bibnamefont {Lacaze}}, \bibinfo {author} {\bibfnamefont {M.}~\bibnamefont {Alba}}, \bibinfo {author} {\bibfnamefont {M.}~\bibnamefont {De~Boissieu}}, \bibinfo {author} {\bibfnamefont {M.}~\bibnamefont {Gailhanou}},\ and\ \bibinfo {author} {\bibfnamefont {M.}~\bibnamefont {Goldmann}},\ }\bibfield  {title} {\bibinfo {title} {Optical gratings formed in thin smectic films frustrated on a single crystalline substrate},\ }\href@noop {} {\bibfield  {journal} {\bibinfo  {journal} {Phys. Rev. E}\ }\textbf {\bibinfo {volume} {70}},\ \bibinfo {pages} {011709} (\bibinfo {year} {2004})}\BibitemShut {NoStop}%
\bibitem [{\citenamefont {Coursault}\ \emph {et~al.}(2016)\citenamefont {Coursault}, \citenamefont {Zappone}, \citenamefont {Coati}, \citenamefont {Boulaoued}, \citenamefont {Pelliser}, \citenamefont {Limagne}, \citenamefont {Boudet}, \citenamefont {Ibrahim}, \citenamefont {De~Martino}, \citenamefont {Alba} \emph {et~al.}}]{coursault2016self}%
  \BibitemOpen
  \bibfield  {author} {\bibinfo {author} {\bibfnamefont {D.}~\bibnamefont {Coursault}}, \bibinfo {author} {\bibfnamefont {B.}~\bibnamefont {Zappone}}, \bibinfo {author} {\bibfnamefont {A.}~\bibnamefont {Coati}}, \bibinfo {author} {\bibfnamefont {A.}~\bibnamefont {Boulaoued}}, \bibinfo {author} {\bibfnamefont {L.}~\bibnamefont {Pelliser}}, \bibinfo {author} {\bibfnamefont {D.}~\bibnamefont {Limagne}}, \bibinfo {author} {\bibfnamefont {N.}~\bibnamefont {Boudet}}, \bibinfo {author} {\bibfnamefont {B.~H.}\ \bibnamefont {Ibrahim}}, \bibinfo {author} {\bibfnamefont {A.}~\bibnamefont {De~Martino}}, \bibinfo {author} {\bibfnamefont {M.}~\bibnamefont {Alba}}, \emph {et~al.},\ }\bibfield  {title} {\bibinfo {title} {Self-organized arrays of dislocations in thin smectic liquid crystal films},\ }\href@noop {} {\bibfield  {journal} {\bibinfo  {journal} {Soft Matter}\ }\textbf {\bibinfo {volume} {12}},\ \bibinfo {pages} {678} (\bibinfo {year} {2016})}\BibitemShut {NoStop}%
\bibitem [{\citenamefont {Leadbetter}\ \emph {et~al.}(1979)\citenamefont {Leadbetter}, \citenamefont {Frost}, \citenamefont {Gaughan}, \citenamefont {Gray},\ and\ \citenamefont {Mosley}}]{leadbetter1979}%
  \BibitemOpen
  \bibfield  {author} {\bibinfo {author} {\bibfnamefont {A.}~\bibnamefont {Leadbetter}}, \bibinfo {author} {\bibfnamefont {J.}~\bibnamefont {Frost}}, \bibinfo {author} {\bibfnamefont {J.}~\bibnamefont {Gaughan}}, \bibinfo {author} {\bibfnamefont {G.}~\bibnamefont {Gray}},\ and\ \bibinfo {author} {\bibfnamefont {A.}~\bibnamefont {Mosley}},\ }\bibfield  {title} {\bibinfo {title} {The structure of smectic {A} phases of compounds with cyano end groups},\ }\href@noop {} {\bibfield  {journal} {\bibinfo  {journal} {J. Phys. (Paris)}\ }\textbf {\bibinfo {volume} {40}},\ \bibinfo {pages} {375} (\bibinfo {year} {1979})}\BibitemShut {NoStop}%
\bibitem [{\citenamefont {Clark}\ \emph {et~al.}(1993)\citenamefont {Clark}, \citenamefont {Bellini}, \citenamefont {Malzbender}, \citenamefont {Thomas}, \citenamefont {Rappaport}, \citenamefont {Muzny}, \citenamefont {Schaefer},\ and\ \citenamefont {Hrubesh}}]{clark_x-ray_1993}%
  \BibitemOpen
  \bibfield  {author} {\bibinfo {author} {\bibfnamefont {N.~A.}\ \bibnamefont {Clark}}, \bibinfo {author} {\bibfnamefont {T.}~\bibnamefont {Bellini}}, \bibinfo {author} {\bibfnamefont {R.~M.}\ \bibnamefont {Malzbender}}, \bibinfo {author} {\bibfnamefont {B.~N.}\ \bibnamefont {Thomas}}, \bibinfo {author} {\bibfnamefont {A.~G.}\ \bibnamefont {Rappaport}}, \bibinfo {author} {\bibfnamefont {C.~D.}\ \bibnamefont {Muzny}}, \bibinfo {author} {\bibfnamefont {D.~W.}\ \bibnamefont {Schaefer}},\ and\ \bibinfo {author} {\bibfnamefont {L.}~\bibnamefont {Hrubesh}},\ }\bibfield  {title} {\bibinfo {title} {X-ray scattering study of smectic ordering in a silica aerogel},\ }\href {https://doi.org/10.1103/PhysRevLett.71.3505} {\bibfield  {journal} {\bibinfo  {journal} {Phys. Rev. Lett.}\ }\textbf {\bibinfo {volume} {71}},\ \bibinfo {pages} {3505} (\bibinfo {year} {1993})}\BibitemShut {NoStop}%
\bibitem [{\citenamefont {Aviles}\ and\ \citenamefont {Giga}(1987)}]{AG}%
  \BibitemOpen
  \bibfield  {author} {\bibinfo {author} {\bibfnamefont {P.}~\bibnamefont {Aviles}}\ and\ \bibinfo {author} {\bibfnamefont {Y.}~\bibnamefont {Giga}},\ }\bibfield  {title} {\bibinfo {title} {A mathematical problem related to the physical theory of liquid crystal configurations},\ }\href@noop {} {\bibfield  {journal} {\bibinfo  {journal} {Proc. Centre Math. Appl.}\ }\textbf {\bibinfo {volume} {1987}},\ \bibinfo {pages} {1} (\bibinfo {year} {1987})}\BibitemShut {NoStop}%
\bibitem [{\citenamefont {Michel}\ \emph {et~al.}(2006)\citenamefont {Michel}, \citenamefont {Lacaze}, \citenamefont {Goldmann}, \citenamefont {Gailhanou}, \citenamefont {De~Boissieu},\ and\ \citenamefont {Alba}}]{michel2006}%
  \BibitemOpen
  \bibfield  {author} {\bibinfo {author} {\bibfnamefont {J.-P.}\ \bibnamefont {Michel}}, \bibinfo {author} {\bibfnamefont {E.}~\bibnamefont {Lacaze}}, \bibinfo {author} {\bibfnamefont {M.}~\bibnamefont {Goldmann}}, \bibinfo {author} {\bibfnamefont {M.}~\bibnamefont {Gailhanou}}, \bibinfo {author} {\bibfnamefont {M.}~\bibnamefont {De~Boissieu}},\ and\ \bibinfo {author} {\bibfnamefont {M.}~\bibnamefont {Alba}},\ }\bibfield  {title} {\bibinfo {title} {Structure of smectic defect cores: X-ray study of 8cb liquid crystal ultrathin films},\ }\href@noop {} {\bibfield  {journal} {\bibinfo  {journal} {Phys. Rev. Lett.}\ }\textbf {\bibinfo {volume} {96}},\ \bibinfo {pages} {027803} (\bibinfo {year} {2006})}\BibitemShut {NoStop}%
\bibitem [{\citenamefont {Blanc}\ and\ \citenamefont {Kleman}(1999)}]{Blanc1999}%
  \BibitemOpen
  \bibfield  {author} {\bibinfo {author} {\bibfnamefont {C.}~\bibnamefont {Blanc}}\ and\ \bibinfo {author} {\bibfnamefont {M.}~\bibnamefont {Kleman}},\ }\bibfield  {title} {\bibinfo {title} {Curvature walls and focal conic domains in a lyotropic lamellar phase},\ }\href@noop {} {\bibfield  {journal} {\bibinfo  {journal} {Eur. Phys. J. B}\ }\textbf {\bibinfo {volume} {10}},\ \bibinfo {pages} {53} (\bibinfo {year} {1999})}\BibitemShut {NoStop}%
\bibitem [{\citenamefont {De~Gennes}\ and\ \citenamefont {Prost}(1993)}]{DeGennes}%
  \BibitemOpen
  \bibfield  {author} {\bibinfo {author} {\bibfnamefont {P.~G.}\ \bibnamefont {De~Gennes}}\ and\ \bibinfo {author} {\bibfnamefont {J.}~\bibnamefont {Prost}},\ }\href@noop {} {\emph {\bibinfo {title} {The Physics of Liquid Crystals}}}\ (\bibinfo  {publisher} {Oxford Science Publications},\ \bibinfo {year} {1993})\BibitemShut {NoStop}%
\bibitem [{\citenamefont {Tintaru}\ \emph {et~al.}(2001)\citenamefont {Tintaru}, \citenamefont {Moldovan}, \citenamefont {Beica},\ and\ \citenamefont {Frunza}}]{tintaru2001}%
  \BibitemOpen
  \bibfield  {author} {\bibinfo {author} {\bibfnamefont {M.}~\bibnamefont {Tintaru}}, \bibinfo {author} {\bibfnamefont {R.}~\bibnamefont {Moldovan}}, \bibinfo {author} {\bibfnamefont {T.}~\bibnamefont {Beica}},\ and\ \bibinfo {author} {\bibfnamefont {S.}~\bibnamefont {Frunza}},\ }\bibfield  {title} {\bibinfo {title} {Surface tension of some liquid crystals in the cyanobiphenyl series},\ }\href@noop {} {\bibfield  {journal} {\bibinfo  {journal} {Liq. Cryst.}\ }\textbf {\bibinfo {volume} {28}},\ \bibinfo {pages} {793} (\bibinfo {year} {2001})}\BibitemShut {NoStop}%
\bibitem [{\citenamefont {Schuring}\ and\ \citenamefont {Stannarius}(2001)}]{schuring2001}%
  \BibitemOpen
  \bibfield  {author} {\bibinfo {author} {\bibfnamefont {C.}~\bibnamefont {Schuring}, \bibfnamefont {H.~Thieme}}\ and\ \bibinfo {author} {\bibfnamefont {R.}~\bibnamefont {Stannarius}},\ }\bibfield  {title} {\bibinfo {title} {Surface tensions of smectic liquid crystals},\ }\href@noop {} {\bibfield  {journal} {\bibinfo  {journal} {Liq. Cryst.}\ }\textbf {\bibinfo {volume} {28}},\ \bibinfo {pages} {241} (\bibinfo {year} {2001})}\BibitemShut {NoStop}%
\bibitem [{\citenamefont {Bradshaw}\ and\ \citenamefont {Raynes}(1985)}]{bradshaw1985}%
  \BibitemOpen
  \bibfield  {author} {\bibinfo {author} {\bibfnamefont {M.~J.}\ \bibnamefont {Bradshaw}}\ and\ \bibinfo {author} {\bibfnamefont {E.~P.}\ \bibnamefont {Raynes}},\ }\bibfield  {title} {\bibinfo {title} {The frank constants of some nematic liquid crystals},\ }\href@noop {} {\bibfield  {journal} {\bibinfo  {journal} {J. Phys. (Paris)}\ }\textbf {\bibinfo {volume} {46}},\ \bibinfo {pages} {1513} (\bibinfo {year} {1985})}\BibitemShut {NoStop}%
\bibitem [{\citenamefont {Benzekri}\ \emph {et~al.}(1992)\citenamefont {Benzekri}, \citenamefont {Claverie}, \citenamefont {Marcerou},\ and\ \citenamefont {Rouillon}}]{Benzekri1992}%
  \BibitemOpen
  \bibfield  {author} {\bibinfo {author} {\bibfnamefont {M.}~\bibnamefont {Benzekri}}, \bibinfo {author} {\bibfnamefont {T.}~\bibnamefont {Claverie}}, \bibinfo {author} {\bibfnamefont {J.}~\bibnamefont {Marcerou}},\ and\ \bibinfo {author} {\bibfnamefont {J.}~\bibnamefont {Rouillon}},\ }\bibfield  {title} {\bibinfo {title} {Nonvanishing of the layer compressional elastic constant at the smetic-a-to-nematic phase transition: A consequence of landau-peierls instability?},\ }\href@noop {} {\bibfield  {journal} {\bibinfo  {journal} {Phys. Rev. Lett.}\ }\textbf {\bibinfo {volume} {68}},\ \bibinfo {pages} {2480} (\bibinfo {year} {1992})}\BibitemShut {NoStop}%
\bibitem [{\citenamefont {Zywocinski}\ \emph {et~al.}(2000)\citenamefont {Zywocinski}, \citenamefont {Picano}, \citenamefont {Oswald},\ and\ \citenamefont {G{\'e}minard}}]{Zywocinski2000}%
  \BibitemOpen
  \bibfield  {author} {\bibinfo {author} {\bibfnamefont {A.}~\bibnamefont {Zywocinski}}, \bibinfo {author} {\bibfnamefont {F.}~\bibnamefont {Picano}}, \bibinfo {author} {\bibfnamefont {P.}~\bibnamefont {Oswald}},\ and\ \bibinfo {author} {\bibfnamefont {J.-C.}\ \bibnamefont {G{\'e}minard}},\ }\bibfield  {title} {\bibinfo {title} {Edge dislocation in a vertical smectic-a film: Line tension versus temperature and film thickness near the nematic phase},\ }\href@noop {} {\bibfield  {journal} {\bibinfo  {journal} {Phys. Rev. E}\ }\textbf {\bibinfo {volume} {62}},\ \bibinfo {pages} {8133} (\bibinfo {year} {2000})}\BibitemShut {NoStop}%
\end{thebibliography}%
\end{document}